\documentclass[aps,prb,reprint,amsfonts,amsmath,amssymb,showpacs,groupedaddress,superscriptaddress]{revtex4-2}
\usepackage[colorlinks,urlcolor=blue,linkcolor=blue,anchorcolor=blue,citecolor=blue,bookmarks]{hyperref}
\usepackage{graphicx}
\usepackage{bm}
\usepackage{color}
\usepackage{float}
\usepackage{subfigure}

\usepackage{xcolor}
\usepackage{ulem}
\usepackage{soul}

\begin{document}

\title{Transition from $s_{\pm}$-wave to $d_{x^{2}-y^{2}}$-wave superconductivity driven by interlayer interaction in the bilayer two-orbital model of La$_3$Ni$_2$O$_7$}
\author{Wenhan Xi}
\affiliation{National Laboratory of Solid State Microstructures and Department of Physics, Nanjing University, Nanjing 210093, China}
\author{Shun-Li Yu}
\email{slyu@nju.edu.cn}
\affiliation{National Laboratory of Solid State Microstructures and Department of Physics, Nanjing University, Nanjing 210093, China}
\affiliation{Collaborative Innovation Center of Advanced Microstructures, Nanjing University, Nanjing 210093, China}
\author{Jian-Xin Li}
\email{jxli@nju.edu.cn}
\affiliation{National Laboratory of Solid State Microstructures and Department of Physics, Nanjing University, Nanjing 210093, China}
\affiliation{Collaborative Innovation Center of Advanced Microstructures, Nanjing University, Nanjing 210093, China}

\date{\today}

\begin{abstract}
    We utilize the fluctuation-exchange approximation on a bilayer two-orbital model, incorporating $d_{x^2-y^2}$ and $d_{z^2}$ orbitals, to explore potential pairing symmetries in the bilayer nickelate La$_3$Ni$_2$O$_7$. Our study particularly examines the impact of interlayer Coulomb interactions. In the absence of these interactions, the superconducting gap exhibits $s_{\pm}$-wave symmetry, with predominant intraorbital pairing in the $d_{z^2}$ orbital. As interlayer interactions increase, $s_{\pm}$-wave superconductivity is suppressed, while the superconductivity with a $d_{x^2-y^2}$-wave gap is enhanced, resulting in a transition at a critical interaction strength. This $d_{x^2-y^2}$-wave superconductivity is distinct not only from the $s_{\pm}$-wave superconductivity but also from the intraorbital $d$-wave pairing in cuprate superconductors, as it is dominated by the interlayer pairing between the $d_{x^2-y^2}$ and $d_{z^2}$ orbitals. Additionally, charge fluctuations play a crucial role in driving the transition from $s_{\pm}$ wave to $d_{x^2-y^2}$ wave superconductivity. Our findings indicate that interlayer Coulomb interactions are crucial for understanding the pairing mechanism in La$_3$Ni$_2$O$_7$.
\end{abstract}

\maketitle

\section{\label{intro}Introduction}

Since the discovery of high-temperature superconductivity in cuprates, the mechanisms underlying this phenomenon and the search for new superconducting materials have remained some of the most challenging and significant topics in condensed matter physics. Recently, the discovery of superconductivity with a transition temperature $T_{c}\approx80~\mathrm{K}$ under high pressure in the nickelate La$_{3}$Ni$_{2}$O$_{7}$~\cite{Sun2023,Nat.Commun.15.2470,Nat.Phys.20.1269,CPL.40.117302,PhysRevX.14.011040,zhou2023evidence,zhang2024147,wang2024normal,li2024pressure} has marked the first instance of achieving high-temperature superconductivity within the liquid nitrogen temperature range in non-cuprate materials. This breakthrough has spurred intensive research activity into Ni-based superconductors.

At ambient pressure, La$_{3}$Ni$_{2}$O$_{7}$ is metallic and exhibits an orthorhombic structure. Upon increasing the pressure, it undergoes a structural transition from the A\textit{mam} to the F\textit{mmm} at approximately 14 GPa~\cite{Sun2023,Wang2024e}, coinciding with the onset of superconductivity. Above $T_{c}$, the resistivity shows a linear temperature dependence~\cite{Sun2023,Nat.Phys.20.1269,Nat.Commun.15.2470}, highlighting the importance of electronic correlations. Although there is no consensus on the precise magnetic properties of this material from experimental studies, techniques such as resonant inelastic X-ray scattering~\cite{Chen2024}, muon-spin relaxation~\cite{PhysRevLett.132.256503,arXiv:2402.10485}, nuclear magnetic resonance~\cite{arXiv:2402.03952,Kakoi2023MultibandMG}, and neutron scattering~\cite{arXiv:2401.12635} all suggest the presence of magnetic ordering or significant magnetic fluctuations. Additionally, a charge density wave may also exist and influence the physical properties of La$_{3}$Ni$_{2}$O$_{7}$~\cite{Nat.Commun.15.7570,Liu2022}. These findings imply that magnetic and charge fluctuations could be crucial in understanding the pairing mechanism in this material.

Since it is not feasible to probe the band structure at high pressure using angle-resolved photoemission spectroscopy, all current theoretical studies on the superconducting mechanism in La$_{3}$Ni$_{2}$O$_{7}$ rely on results from density functional theory (DFT) calculations~\cite{Sun2023,Christiansson2023,Sui2024,PhysRevB.109.L081105,PhysRevLett.131.126001,PhysRevB.108.L180510,PhysRevB.108.L201121,arXiv:2401.15097}. DFT results under high pressure reveal that the low energy bands near the Fermi level are predominantly composed of Ni-$3d_{x^2-y^2}$ and Ni-$3d_{z^2}$ orbitals. This has consequently led to the widespread adoption of the bilayer two-orbital model in theoretical studies~\cite{PhysRevLett.131.126001,PhysRevB.108.L180510,PhysRevB.108.L201121,arXiv:2401.15097}. However, some researches have proposed that the band structure of La$_{3}$Ni$_{2}$O$_{7}$ is analogous to that of the bilayer cuprates, leading to the use of a bilayer single-orbital model as well~\cite{PhysRevB.110.024514}. Additionally, certain studies also suggest that in the strong coupling limit, the two-orbital model can be approximated as a single-orbital model~\cite{PhysRevLett.132.146002}.

A variety of methods have been used to explore the nature of superconducting pairing symmetry, including the functional renormalization group (fRG), tensor network calculations, random phase approximation (RPA), and dynamical mean-field theory (DMFT)~\cite{PhysRevB.108.L140505,Qin2023,Yang2023a,PhysRevLett.131.236002,PhysRevB.108.165141,PhysRevB.108.214522,arXiv:2306.07275,PhysRevLett.132.106002,PhysRevLett.132.036502,
PhysRevB.109.165154,arXiv:2312.03605,npjQuantumMater.9.61,PhysRevLett.133.096002,arXiv:2409.17861,
PhysRevB.108.L201121,arXiv:2401.15097,PhysRevB.110.024514,PhysRevLett.132.146002,PhysRevB.108.174501,CPL.41.017402,
PhysRevB.109.104508,PhysRevLett.132.126503,ShenShenYang2023,Xia2025}. Theoretical results vary significantly based on the models and methods applied. In studies using a two-orbital model with $d_{x^2-y^2}$ and $d_{z^2}$ orbitals, fRG~\cite{PhysRevB.108.L140505}, RPA~\cite{PhysRevLett.131.236002,Xia2025} and DMFT~\cite{PhysRevB.109.165154,PhysRevLett.133.096002} studies suggested an $s_{\pm}$-wave symmetry driven by inter-pocket spin fluctuations. One study also revealed that minor adjustments to the crystal field splitting ($\sim 0.2$ eV) can switch the dominant symmetry between $s_{\pm}$ to $d_{xy}$~\cite{Xia2025}, highlighting the sensitivity of pairing symmetry to low-energy electronic structures. Conversely, bilayer single-orbital models predominantly predict $d$-wave pairing mediated by antiferromagnetic spin fluctuations~\cite{PhysRevB.110.024514}. A slave-boson mean-field analysis of the bilayer single-orbital $t$-$J$-$J_{\perp}$ models revealed that strong interlayer magnetic exchange $J_{\perp}$  tunes the single-layer $d$-wave superconducting state to the $s$-wave one with dominant interlayer pairings~\cite{PhysRevLett.132.146002}, a result also supported by tensor network calculations~\cite{PhysRevLett.132.036502}.

Although there is currently no consensus on the pairing symmetry of La$_{3}$Ni$_{2}$O$_{7}$ in theoretical researches, with both $s$-wave and $d$-wave pairing symmetries being proposed, almost all studies indicate that the interlayer coupling of the Ni-$3d_{z^2}$ orbital plays a crucial role in determining the pairing symmetry. Given the strong interlayer coupling of the Ni-$3d_{z^2}$ orbital, it naturally leads us to ponder whether the interlayer Coulomb interactions of the $3d_{z^2}$ orbital could also play a significant role in determining the pairing symmetry. While some studies have discussed the impact of interlayer superexchange on superconductivity~\cite{PhysRevLett.132.146002,Qin2023,PhysRevLett.132.036502}, most previous research has focused primarily on intralayer Coulomb interactions, often neglecting the interlayer aspects. Therefore, analyzing the influence of interlayer Coulomb interactions on pairing symmetry may be crucial for revealing the superconducting pairing mechanism of La$_{3}$Ni$_{2}$O$_{7}$.

In this paper, we employ a bilayer two-orbital model consisting of $d_{x^2-y^2}$ and $d_{z^2}$ orbitals, using the fluctuation-exchange (FLEX) approximation to study the potential superconducting pairing symmetries of La$_{3}$Ni$_{2}$O$_{7}$. We pay particular attention to the effects of the interlayer Coulomb interaction $V_{z}$ between the $d_{z^2}$ orbitals. For $V_z=0$, we find that the superconducting gap exhibits $s_{\pm}$-wave symmetry, with dominant intraorbital pairing of the $d_{z^2}$ orbital and similar amplitudes for both intralayer and interlayer pairings. As $V_z$ increases, the $s_{\pm}$-wave superconductivity is gradually suppressed, while the $d_{x^2-y^2}$-wave superconductivity is enhanced, leading to a transition to the $d_{x^2-y^2}$-wave superconductivity at a critical $V_z$. During this process, charge fluctuations associated with $V_z$ play a critical role. Moreover, unlike the $s_{\pm}$-wave superconductivity, the $d_{x^2-y^2}$-wave superconductivity is dominated by the interlayer pairing between the $d_{x^2-y^2}$ and $d_{z^2}$ orbitals. Furthermore, due to the mirror symmetry of the bilayer structure, pairing symmetries can also be described using the bonding-antibonding basis. In this framework, the $s_{\pm}$-wave superconductivity is dominated by intraorbital pairing of the antibonding $d_{z^2}$ orbital, while the $d_{x^2-y^2}$-wave superconductivity is dominated by interorbital pairing between the bonding $d_{x^2-y^2}$ and $d_{z^2}$ orbitals.

The paper is organized as follows: In section~\ref{model-method}, we introduce the model and the FLEX method, highlighting some unique properties of the model. Section~\ref{result} details the evolution of superconducting pairing symmetries and thoroughly discusses the underlying mechanisms. Section~\ref{summary} presents a summary.

\section{\label{model-method}Model and Method}

Based on DFT calculations predicting dominant Ni-$3d_{x^2-y^2}$ and Ni-$3d_{z^2}$ characteristics near the Fermi level~\cite{PhysRevLett.131.126001,PhysRevB.108.L180510,PhysRevB.108.L201121,arXiv:2401.15097,arXiv:2306.07275}, we employ the following effective bilayer two-orbital tight-binding (TB) model~\cite{PhysRevLett.131.126001}:
\begin{align}
H_0 =\sum_{\substack{i\delta,ll^{\prime}\\ aa^{\prime},\sigma}}t^{\delta}_{ll',aa'}c^\dagger_{ila\sigma}c_{i+\delta,l'a'\sigma}
    - \sum_{ila\sigma}(\varepsilon_{a}+\mu)n_{ila\sigma},
\label{TB-model}
\end{align}
where $c_{ila\sigma}$ is the electron annihilation operator associated with site $i$, layer $l=\{A,B\}$, orbital $a=\{x,z\}$ and spin $\sigma$. $n_{ila\sigma}=c^\dagger_{ila\sigma}c_{ila\sigma}$ is the particle number operator. The layers are labeled $A$ and $B$, while the $d_{x^2-y^2}$ and $d_{z^2}$ orbitals are denoted by $x$ and $z$. The parameter $\delta=\{1,2\}$ represents the nearest-neighbor (NN) and next-nearest-neighbor (NNN) hoppings, and $\mu$ is the chemical potential. We adopt the TB parameters from Luo et al. in Ref.~\onlinecite{PhysRevLett.131.126001}. The onsite energies are $(\varepsilon_{x}, \varepsilon_{z})=(-0.776,-0.409)$ eV. The intralayer hoppings are $(t^{1}_{ll,xx},t^{2}_{ll,xx},t^{1}_{ll,zz},t^{2}_{ll,zz},t^{1}_{ll,xz})=(-0.483,0.069,-0.11,-0.017,0.239)$ eV, and the interlayer hoppings are $(t^{1}_{ll^{\prime},xz},t^{1}_{ll^{\prime},xx},t^{1}_{ll^{\prime},zz})=(-0.034,0.005,-0.635)$ eV for $l\neq l^{\prime}$. The significant interlayer hopping of the $d_{z^2}$ orbital and the strong intralayer hybridization between different orbitals indicate that neither the two orbitals nor the two layers can be simply decoupled. These are key features of the TB model.

We can transform the Hamiltonian (\ref{TB-model}) into momentum space as follows:
\begin{align}
H_0=\sum_{\bm{k}\sigma}\Psi^{\dag}_{\bm{k}\sigma}M(\bm{k})\Psi_{\bm{k}\sigma},
\label{TB-momentum}
\end{align}
where $\Psi_{\bm{k}\sigma}=(c_{kAx\sigma},c_{kAz\sigma},c_{kBx\sigma},c_{kBz\sigma})^{\mathrm{T}}$, and $M(\bm{k})$ is a $4\times4$ Hermitian matrix. Here, $c_{kla\sigma}$ is the Fourier transformation from $c_{ila\sigma}$, and the elements of $M(\bm{k})$ are derived from the Fourier transformation of the Hamiltonian (\ref{TB-model}).

\begin{figure}
    \centering
    \includegraphics[width=\columnwidth]{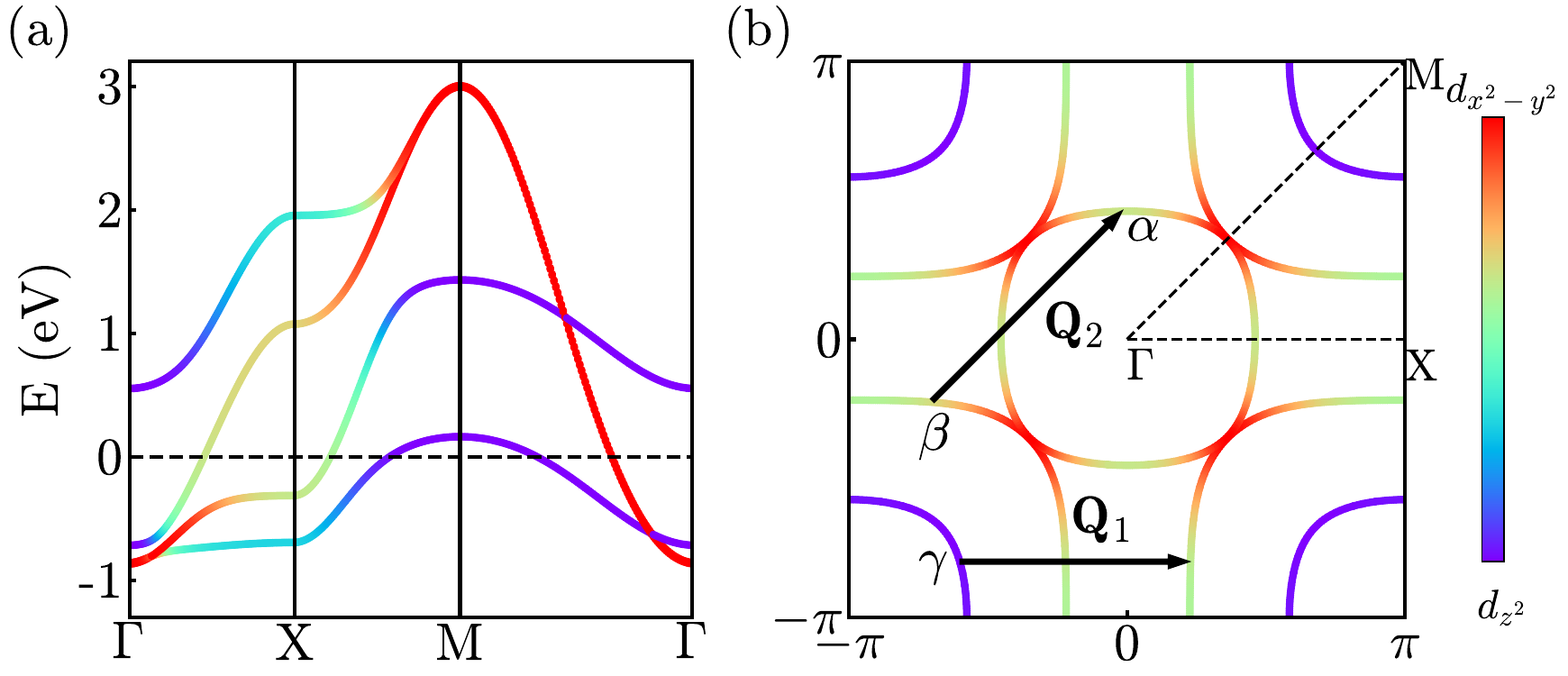}
    \caption{Band structure (a) and Fermi surface (b) of  the bilayer two-orbital model. Colors indicate orbital weights. The three Fermi pockets in (b) are labeled $\alpha$, $\beta$, and $\gamma$. Black arrows $\bm{Q}_1$ and $\bm{Q}_2$ in (b) denote the nesting wave vectors.}
    \label{FS_band}
\end{figure}

The band structure and Fermi surface (FS) of the model (\ref{TB-model}) with chemical potential $\mu=0$ are shown in Fig.~\ref{FS_band}(a) and (b), respectively. This choice of chemical potential corresponds to the nominal $3d^{7.5}$ electronic configuration in  La$_{3}$Ni$_{2}$O$_{7}$~\cite{Sun2023}, where the three $t_{2g}$ orbitals are fully filled and the two $e_g$ orbitals have an average occupancy of 1.5 electrons. In this bilayer two-orbital model, this is equivalent to an average filling of 3 electrons per site. There are three sets of FS: an electron pocket $\alpha$ near the $\Gamma$ point, and a hole pocket $\beta$ and a hole pocket $\gamma$ around the M point. The small $\gamma$ pocket is predominantly composed of $d_{z^2}$ orbital, whereas the $\alpha$ and $\beta$ pockets arise from the hybridization of $d_{x^2-y^2}$ and $d_{z^2}$ orbitals. Additionally, the FS depicted in Fig.~\ref{FS_band}(b) features two distinct sets of nesting wave vectors, denoted by $\bm{Q}_1$ and $\bm{Q}_2$, which are expected to lead to strong spin and charge fluctuations.

Due to the mirror symmetry of the bilayer structure, the matrix $M(\bm{k})$ in Eq.~(\ref{TB-momentum}) can be expressed as a block diagonal matrix in the bonding-antibonding (BA) basis:
\begin{align}
H_0=\sum_{\bm{k}\sigma}\widetilde{\Psi}^{\dagger}_{\bm{k}\sigma}\widetilde{M}(\bm{k})\widetilde{\Psi}_{\bm{k}\sigma}.
\label{TB-momentum-AB}
\end{align}
Here, $\widetilde{\Psi}_{\bm{k}\sigma}=(c_{k,+,x\sigma},c_{k,+,z\sigma},c_{k,-,x\sigma},c_{k,-,z\sigma})^{\mathrm{T}}$, where $c_{k,\pm,a\sigma} = (c_{kAa\sigma} \pm c_{kBa\sigma})/\sqrt{2}$ represents the bonding and antibonding states. The matrix $\widetilde{M}(\bm{k})$ is given by
\begin{align}
\widetilde{M}(\bm{k})=U^{\dag}M(\bm{k})U,
\end{align}
where
\begin{align}
U=\frac{1}{\sqrt{2}}\left(
                      \begin{array}{cccc}
                        1 & 0 & 1 & 0 \\
                        0 & 1 & 0 & 1 \\
                        1 & 0 & -1 & 0 \\
                        0 & 1 & 0 & -1 \\
                      \end{array}
                    \right)
\label{U-O-BA}
\end{align}
is the transformation matrix from the orbital-layer representation to the BA representation. Since there is no mixing between the bonding and antibonding states, the $\alpha$ and $\gamma $ pockets are derived entirely from the bonding states, while the $\beta$ pocket is derived entirely from the antibonding states. As illustrated in Fig.~\ref{FS_band}, the bonding state of the $d_{z^2}$ orbital is predominantly on the $\gamma$ pocket, whereas the antibonding state of the $d_{z^2}$ orbital is mainly on the $\beta$ pocket. The bonding state of the $d_{x^2-y^2}$ orbital is confined to the $\alpha$ pocket,  while its antibonding state is located on the $\beta$ pocket. These distributions of the bonding and antibonding orbitals impose significant constraints on the superconducting pairing symmetry, as we will discuss below.

The strong interlayer hopping of the $d_{z^2}$ orbital suggests that interlayer superconducting pairing in this orbital is also likely to be significant, potentially making interlayer interactions crucial for determining the pairing symmetry. However, previous theoretical studies often overlooked these interlayer interactions~\cite{PhysRevB.108.L201121,arXiv:2401.15097,arXiv:2306.07275,PhysRevB.110.024514,PhysRevLett.132.146002,PhysRevB.108.L140505,
PhysRevLett.132.106002,PhysRevLett.131.236002,Nat.Commun.15.2470,PhysRevLett.132.036502,PhysRevB.108.174501,
PhysRevB.108.165141,PhysRevB.109.165154,arXiv:2312.03605,PhysRevB.108.214522,CPL.41.017402,
PhysRevB.109.104508,npjQuantumMater.9.61,PhysRevLett.133.096002,PhysRevLett.132.126503}. By considering both intralayer and interlayer interactions, we derive the following interaction form:
\begin{align}
H_{int} = H_{intra}+H_{inter},
\end{align}
where
\begin{align}
    H_{intra} =&  \frac{U}{2}\sum_{ila,\sigma \neq \sigma'} n_{ila\sigma}n_{ila\sigma'}+\frac{U'}{2}\sum_{il,\sigma\sigma',a \neq a'}n_{ila\sigma}n_{ila'\sigma'} \nonumber \\
    &+\frac{J}{2} \sum_{il,a \neq a',\sigma\sigma'} c^{\dagger}_{ila\sigma}c^{\dagger}_{ila'\sigma'}c_{ila\sigma'}c_{ila'\sigma} \nonumber \\
   &+\frac{J'}{2}\sum_{il,a \neq a',\sigma \neq \sigma'}c^\dagger_{ila\sigma}c^\dagger_{ila\sigma'}c_{ila'\sigma'}c_{ila'\sigma}
\end{align}
and
\begin{align}
     H_{inter} =  \frac{V_{z}}{2}\sum_{i,l \neq l',\sigma\sigma'}n_{ilz\sigma}n_{il'z\sigma'}.
\end{align}
Here, $U$ ($U'$) is the intraorbital (interorbital) Coulomb interaction, $J$ is the Hund's coupling, and $J'$ is the interorbital pairing hopping, all within a layer. $V_z$ denotes the interlayer Coulomb interaction of the $d_{z^2}$ orbital. In our calculations, the relations $U=U'+2J$ and $J=J'$ are used.

Given recent experiments suggesting the possible presence of spin and charge density waves in La$_{3}$Ni$_{2}$O$_{7}$ at ambient pressure~\cite{Chen2024,PhysRevLett.132.256503,arXiv:2402.10485,arXiv:2402.03952,arXiv:2401.12635,Nat.Commun.15.7570}, spin and charge fluctuations are likely to play a crucial role in superconducting pairing under high pressure. To explore this, we employ the FLEX approximation~\cite{bickers1989conserving2,spin2013yu,wang2015fermi,zijianyao2009}, a self-consistent conserving method that is particularly effective for studying superconductivity mediated by various collective fluctuations. For the bilayer two-orbital model, the Green's function and self-energy are expressed as $4\times4$ matrices, satisfying:
\begin{align}
G^{-1}(k)=G_{0}^{-1}(k)-\Sigma(k)
\end{align}
and
\begin{align}
\Sigma_{m n}(k)=\frac{T}{N} \sum_{q,uv} V_{n u, m v}(q) G_{u v}(k-q),
\end{align}
where each subscript combines layer ($l$) and orbital ($a$) indices, e.g., $m=(l_m,a_m)$. Here, $T$ represents the temperature, $k=(\bm{k},i\omega_n)$ with Matsubara frequency $\omega_{n}=(2n+1)\pi T$, and $N$ is the total number of lattice sites. The bare
Green's function is given by $G_{0}(k)=[i\omega_{n}-M(\bm{k})]^{-1}$. The interaction vertex $V$ is represented as a $16\times16$ matrix:
\begin{align}
V(q)=&\frac{3}{2} U^{s}\left[\chi^{s}(q)-\chi^{0}(q)\right]U^{s}
\!+\frac{1}{2} U^{c}\left[\chi^{c}(q)-\chi^{0}(q)\right] U^{c} \nonumber \\
&+\frac{1}{2}\left[U^{s} \chi^{0}(q) U^{s}\!+U^{c} \chi^{0}(q) U^{c}\right]
\end{align}
with spin susceptibility $\chi^{s}(q)=[1-\chi^{0}(q) U^{s}]^{-1}\chi^{0}(q)$ and charge susceptibility $\chi^{c}(q)=[1+\chi^{0}(q) U^{c}]^{-1}\chi^{0}(q)$. The irreducible susceptibility is given by
\begin{align}
\chi^{0}_{mn,uv}(q)=-\frac{T}{N}\sum_{k}G_{um}(k+q)G_{nv}(k).
\end{align}
The matrices $U^{s}$ and $U^{c}$ are defined as follows: for $a_{m}=a_{n}=a_{u}=a_{v}$ and $l_{m}=l_{n}=l_{u}=l_{v}$, $U^{s}_{mn,uv}=U$, $U^{c}_{mn,uv}=U$; for $a_{m}=a_{u}\neq a_{n}=a_{v}$ and $l_{m}=l_{n}=l_{u}=l_{v}$, $U^{s}_{mn,uv}=U'$, $U^{c}_{mn,uv}=2J-U'$; for $a_{m}=a_{n}\neq a_{u}=a_{v}$ and $l_{m}=l_{n}=l_{u}=l_{v}$, $U^{s}_{mn,uv}=J$, $U^{c}_{mn,uv}=2U'-J$; for $a_{m}=a_{v}\neq a_{n}=a_{u}$ and $l_{m}=l_{n}=l_{u}=l_{v}$, $U^{s}_{mn,uv}=J'$, $U^{c}_{mn,uv}=J'$; for $a_{m}=a_{u}=a_{n}=a_{v}=d_{z^2}$ and $l_{m}=l_{u}\neq l_{n}=l_{v}$, $U^{s}_{mn,uv}=V_z$, $U^{c}_{mn,uv}=-V_z$; for $a_{m}=a_{u}= a_{n}=a_{v}=d_{z^2}$ and $l_{m}=l_{n}\neq l_{u}=l_{v}$,
$U^{s}_{mn,uv}=0$, $U^{c}_{mn,uv}=2V_z$; for other cases, $U^{s}_{mn,uv}=0$, $U^{c}_{mn,uv}=0$.
These equations are solved self-consistently with $N=64\times64$ $\bm{k}$-point meshes and $1024$ $\omega_{n}$.

Assuming that the pairing interaction responsible for superconductivity arises from the exchange of spin and charge fluctuations, we can derive the effective pairing interaction using the FLEX approximation~\cite{bickers1989conserving2,spin2013yu,wang2015fermi}. The singlet pairing interaction is given by
\begin{align}
\Gamma(q)=&\Gamma^{s}(q)+\Gamma^{c}(q)+\Gamma^{0}(q) \nonumber \\
=&\frac{3}{2}U^{s}\chi^{s}(q)U^{s}\!-\!\frac{1}{2}U^{c}\chi^{c}(q)U^{c}\!+\!\frac{1}{2}(U^{s}+U^{c}).
\label{gamma-q-eqx}
\end{align}
In this expression, $\Gamma^{s}$, $\Gamma^{c}$ and $\Gamma^{0}$ represent the contributions from spin fluctuations, charge fluctuations, and the zero-order term, respectively. The superconducting pairing functions can be obtained by solving the Eliashberg equation:
\begin{align}
    \lambda\phi_{ij}(k)=-\frac{T}{N}\!\!\sum_{p,mnuv}\!\!\!\Gamma_{mijn}(k-p)G_{m u}(p)G_{n v}(-p)\phi_{uv}(p).
    \label{gap-eq}
\end{align}
The eigenvector $\phi(k)$ with the largest eigenvalue $\lambda$ indicates the most favorable superconducting pairing symmetry.

The symmetries of the $d_{x^2-y^2}$ and $d_{z^2}$ orbitals impose a specific constraint on the pairing functions. Under reflection across the line $y=x$ (or $y=-x$), the matrix $M(\bm{k})$ in Hamiltonian (\ref{TB-momentum}) satisfies the relation:
\begin{align}
    M(\hat{R}\bm{k}) = U^{\dag}_R M(\bm{k}) U_R,
\end{align}
where $\hat{R}$ represents the reflection across the line $y=x$ and $U_R=\text{diag}(1, -1, 1, -1)$. The bands, obtained by diagonalizing $M(\bm{k})$ with a unitary transformation $U_{b}(\bm{k})$, remain invariant under the reflection $\hat{R}$. Thus, the matrix $U_{b}(\bm{k})$ satisfies $U_{b}(\hat{R} \bm{k}) = U_R U_{b}(\bm{k})$. The gap function $\Delta(\bm{k})$ in the band representation, obtained from $\phi(\bm{k})$ in the orbital-layer representation using the matrix $U_{b}(\bm{k})$ via $\Delta(\bm{k})=U^{\mathrm{T}}_{b}(\bm{k})\phi(\bm{k})U_{b}(\bm{k})$, must satisfy $\Delta(\hat{R}\bm{k})=\pm\Delta(\bm{k})$. Consequently, the pairing function $\phi(k)$ satisfies:
\begin{align}
\phi(\hat{R} \bm{k}) = \pm U_R \phi(\bm{k}) U_R.
\label{eq:constraint}
\end{align}
This imposes a strong constraint on the pairing function: if the intraorbital pairings do not change sign under the transformation $\hat{R}$, then the interorbital pairings must change sign, and vice versa. This constraint applies to both the orbital-layer and BA representations.

\section{\label{result}results and discussion}

In this section, we explore how pairing symmetry evolves with changes in interaction parameters. For convenience, in the orbital-layer representation, the orbital and layer indices $(Ax,Az,Bx,Bz)$ are denoted as $(1,2,3,4)$. In the BA representation, the indices $(x+,z+,x-,z-)$ are denoted as $(x,z,\bar{x},\bar{z})$.

\subsection{The case for $V_z=0$}

\begin{figure}
    \centering
    \includegraphics[width=\columnwidth]{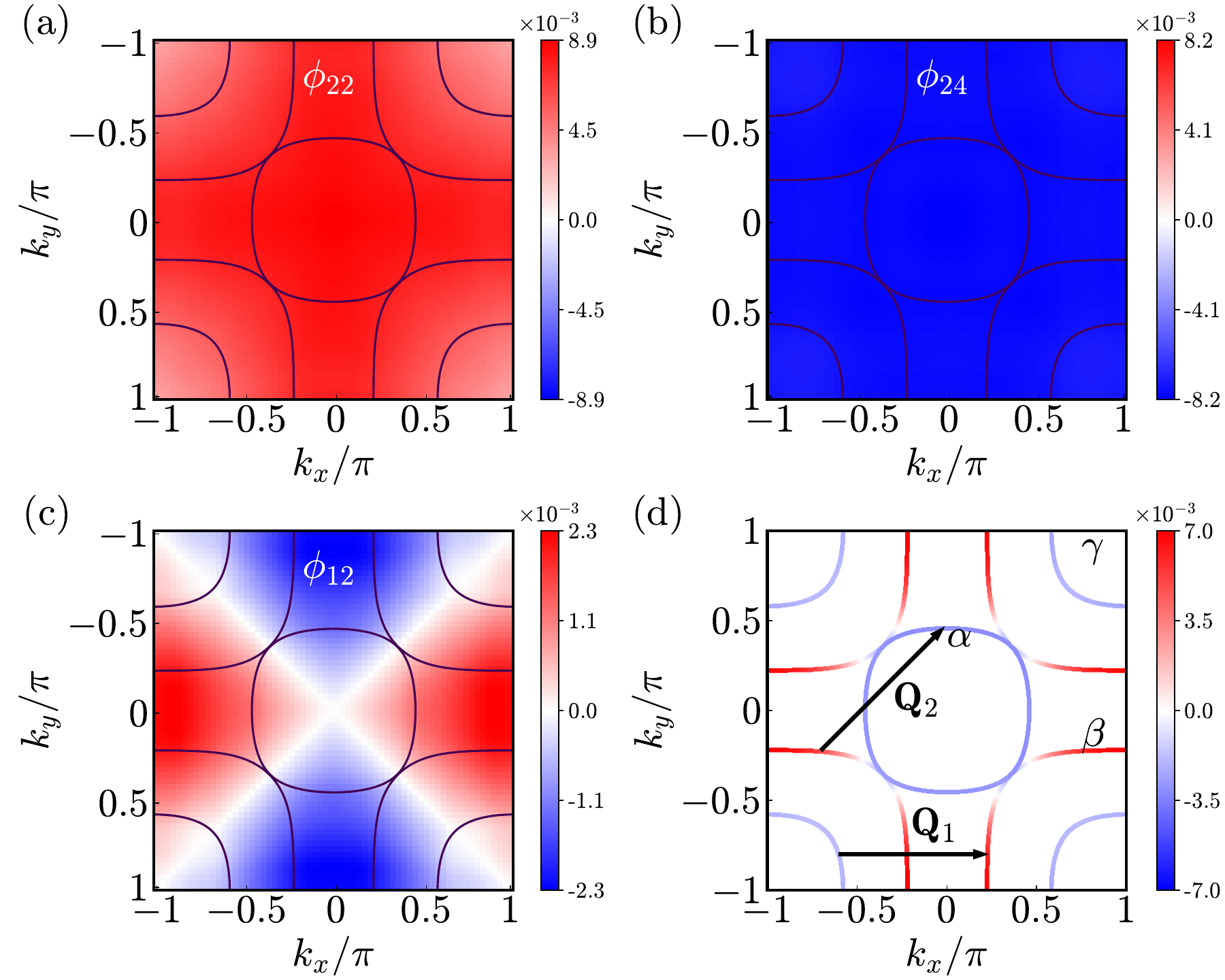}
    \caption{Pairing functions for $U=1.0$ eV and $V_z=0$ in the orbital-layer representation. (a) Intralayer pairing $\phi_{22}$ for the $d_{z^{2}}$ orbital. (b) Interlayer pairing $\phi_{24}$ for the $d_{z^{2}}$ orbital. (c) Intralayer pairing $\phi_{12}$ between the $d_{x^{2}-y^{2}}$ and $d_{z^{2}}$ orbitals. (d) Gap Function projected onto the Fermi surface, with black arrows $\boldsymbol{Q}_1$ and $\boldsymbol{Q}_2$ indicating sign changes.}
    \label{layer_s}
\end{figure}

Let's begin with the situation where there are no interlayer interactions, i.e., $V_z=0$, and set $U=1.0$ eV. We have verified that the results for $U=0.5$ eV and $U=1.5$ eV are qualitatively consistent with those for $U=1.0$ eV. The chemical potential is set to $\mu=0$, corresponding to the case of La$_{3}$Ni$_{2}$O$_{7}$, with an average electron density per site of $n=3$ in the bilayer two-orbital model (\ref{TB-model}). In Fig.~\ref{layer_s}(a)-(c), we present the three main components of the pairing function $\phi(\bm{k})$. It is evident that the dominant pairings are the intralayer pairing $\phi_{22}$ and the interlayer pairing $\phi_{24}$ of the $d_{z^2}$ orbitals, both having nearly the same magnitude. These pairings maintain their signs throughout the entire Brillouin zone (BZ), exhibiting $s$-wave characteristics, but with opposite signs. In contrast, as shown in Fig.~\ref{layer_s}(c), the interorbital pairing between the $d_{x^2-y^2}$ and $d_{z^2}$ orbitals is dominated by the intralayer component $\phi_{12}$ and exhibits $d_{x^{2}-y^{2}}$-wave characteristics. These features of the pairing function satisfy the constraint condition (\ref{eq:constraint}) well. The pairing function can be explicitly expressed as:
\begin{align}
\phi_{ij}(\bm{k})&=
\left\{
\begin{array}{ll}
    \Delta_{\alpha} &  ij=22~\text{or}~ 44  \\
    \Delta_{\beta} &  ij=24 \\
    \Delta_{\gamma}(\cos k_x-\cos k_y) &  ij=12~\text{or}~34 \\
    0 & \text{otherwise}
\end{array}
\right.
\label{gap-spm-x1}
\end{align}
with $\Delta_{\alpha}:\Delta_{\beta}:\Delta_{\gamma}=1:-1.11:0.17$ and $\phi_{ij}(\bm{k})=\phi_{ji}(\bm{k})$. Upon projecting the gap function onto the FS, as illustrated in Fig.~\ref{layer_s}(d), it is observed that the gap functions on the $\alpha$ and $\gamma$ pockets are essentially opposite in sign compared to that on the $\beta$ pocket, indicating characteristics of $s_{\pm}$-wave symmetry, consistent with the other recent theoretical studies, including fRG, RPA and DMFT approaches using similar bilayer two-orbital models~\cite{arXiv:2306.07275,Qin2023,Yang2023a,PhysRevLett.131.236002,PhysRevB.108.L140505,PhysRevLett.132.106002,PhysRevLett.132.036502,
PhysRevB.108.165141,PhysRevB.109.165154,arXiv:2312.03605,PhysRevB.108.214522,npjQuantumMater.9.61,PhysRevLett.133.096002,arXiv:2409.17861}. These characteristics also suggest that the superconducting pairing is predominantly in the $d_{z^2}$ orbital and exhibits strong interlayer pairing. This is a significant departure from results obtained from bilayer single-orbital models~\cite{PhysRevLett.132.146002,PhysRevLett.132.036502}, where superconducting pairing mainly occurs in the $d_{x^{2}-y^{2}}$ orbitals.

To gain a more comprehensive understanding of the origin of the $s_{\pm}$-wave superconducting gap shown in Fig.~\ref{layer_s}(d), we utilize the BA representation. The mirror symmetry ensures that there is no mixing between the bonding and antibonding states. Specifically, the $\alpha$ and $\gamma$ pockets are composed of bonding states, while the $\beta$ pocket arises from antibonding states. For Cooper pairs with zero center-of-mass momentum, there is no pairing between bonding and antibonding states. This separation is highly advantageous for analyzing the structure of the gap function.

\begin{figure}
    \centering
    \includegraphics[width=\columnwidth]{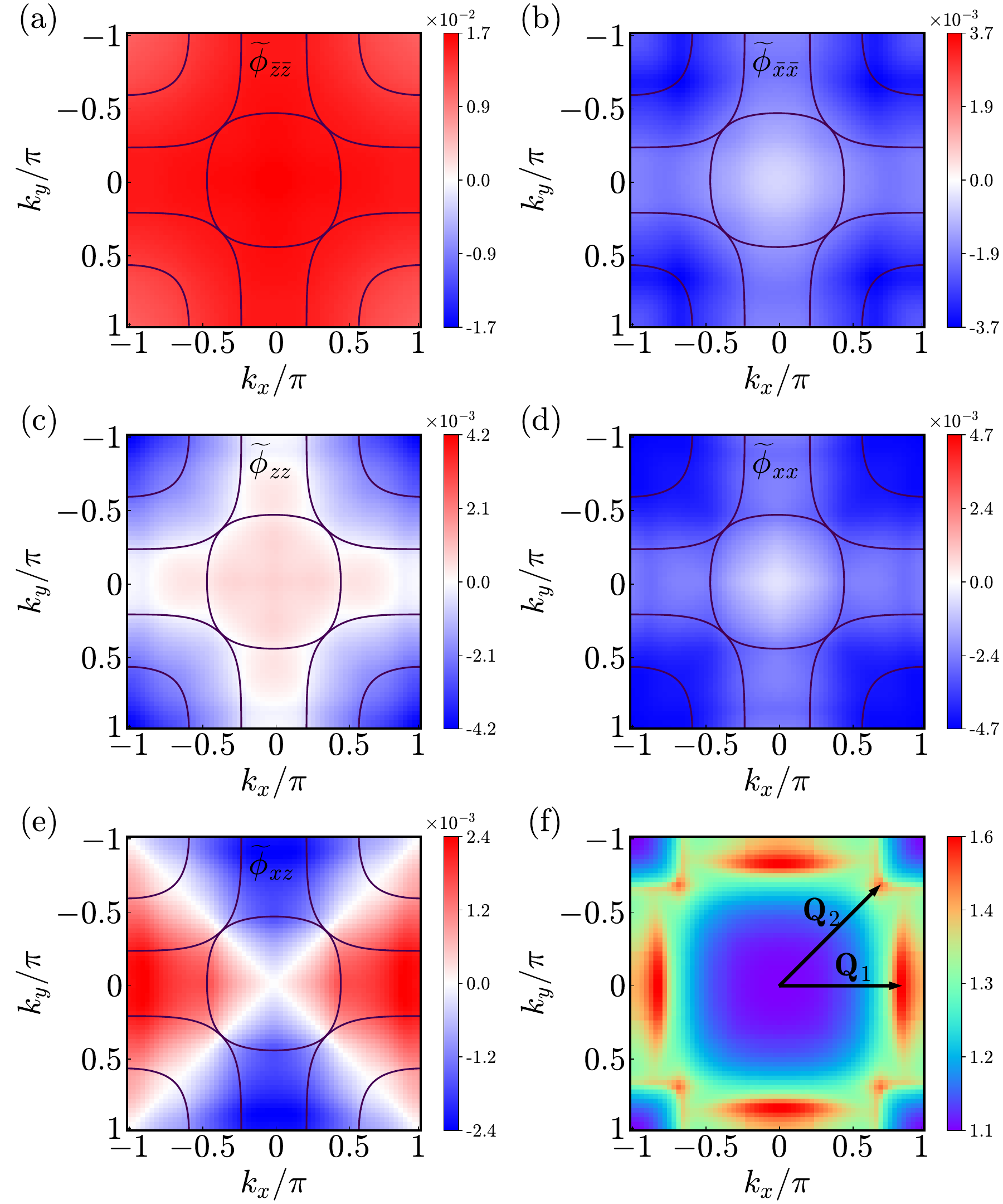}
    \caption{Pairing functions and pairing interaction in the bonding-antibonding representation for $U=1.0$ eV and $V_z=0$. (a)-(e) Pairing functions. (f) Pairing interaction $\widetilde{\Gamma}_{z\bar{z}\bar{z}z}(\bm{k}$), with black arrows $\bm{Q}_1$ and $\bm{Q}_2$ indicating peaks.}
    \label{bond_s}
\end{figure}

As illustrated in Fig.~\ref{bond_s}(a)-(e), within the BA representation,  the pairing functions adhere to constraint (\ref{eq:constraint}), just as in the orbital-layer representation [Fig.~\ref{layer_s}(a)-(c)]. The pairing function $\widetilde{\phi}_{\bar{z}\bar{z}}$ [Fig.~\ref{bond_s}(a)], associated with the antibonding state of the $d_{z^{2}}$ orbital, exhibits the largest pairing magnitude, significantly exceeding other pairing components. Since the antibonding state of the $d_{z^{2}}$ orbital contributes solely to the $\beta$ pocket,  this results in the superconducting gap on the $\beta$ pocket having the maximal value [Fig.~\ref{layer_s}(d)]. We also observed that the superconducting gap on the $\beta$ pocket is not uniform and contains nodes, whereas the pairing function $\widetilde{\phi}_{\bar{z}\bar{z}}$ is almost uniform throughout the BZ. This discrepancy arises due to the contribution of the antibonding state of the $d_{x^{2}-y^{2}}$ orbital. As shown in Fig.~\ref{bond_s}(b), the pairing function $\widetilde{\phi}_{\bar{x}\bar{x}}$, associated with the antibonding state of the $d_{x^{2}-y^{2}}$ orbital, exhibits an opposite sign to $\widetilde{\phi}_{\bar{z}\bar{z}}$. Although $\widetilde{\phi}_{\bar{x}\bar{x}}$ is much smaller than $\widetilde{\phi}_{\bar{z}\bar{z}}$, the weight of the $d_{x^{2}-y^{2}}$ orbital is significantly greater than that of the $d_{z^{2}}$ orbital in the $\Gamma$-$M$ direction [Fig.~\ref{FS_band}(b)], leading to a change in sign of the superconducting gap on the $\beta$ pocket and consequently the presence of a node.

Another important feature of the gap function, as shown in Fig.~\ref{layer_s}(d), is the sign change between the bonding pockets ($\alpha$ and $\gamma$) and the antibonding pocket ($\beta$), indicated by the vectors $\bm{Q}_{1}$ and $\bm{Q}_{2}$. Since there is no pairing between bonding and antibonding states, analyzing the sign changes of the gap function under the BA representation becomes much simpler. The pairing equation in the BA representation can be derived from equation (\ref{gap-eq}) using the transformation (\ref{U-O-BA}). For the bonding and antibonding states of the $d_{z^{2}}$ orbital, the pairing functions satisfy the following relation:
\begin{align}
\lambda \widetilde{\phi}_{\bar{z}\bar{z}}(\bm{k}) \sim& -\frac{T}{N}\sum_{\bm{q}} \widetilde{\Gamma}_{z\bar{z}\bar{z}z}(\bm{q})\widetilde{G}_{zz}(\bm{k}-\bm{q}) \widetilde{G}_{zz}(\bm{q}-\bm{k}) \nonumber \\
&\times \widetilde{\phi}_{zz}(\bm{k}-\bm{q}).
\label{eq-x1}
\end{align}
As shown in Fig.~\ref{bond_s}(f), the pairing interaction $\widetilde{\Gamma}_{z\bar{z}\bar{z}z}$ peaks at the wave vector $\bm{Q}_{1}$, originating from enhanced spin and charge fluctuations due to FS nesting between the $\beta$ and $\gamma$ pockets [see Fig.~\ref{FS_band}(b)]. This means that the pairing scatterings are dominated by transitions from $(\bm{k},-\bm{k})$ to $(\bm{k}+\bm{Q}_{1},-\bm{k}-\bm{Q}_{1})$. To achieve a maximal $\lambda$, the pairing function must satisfy the condition $\widetilde{\phi}_{\bar{z}\bar{z}}(\bm{k})\widetilde{\phi}_{zz}(\bm{k}-\bm{Q}_{1})<0$ [see Eq.~(\ref{eq-x1})]. This results in the bonding and antibonding states of the $d_{z^{2}}$ orbital, connected by $\bm{Q}_{1}$ [Fig.~\ref{layer_s}(d)], having opposite signs [Figs.~\ref{bond_s}(a) and (c)]. Consequently, the gap functions on the $\beta$ and $\gamma$ pockets also exhibit opposite signs. In the BA representation, $\widetilde{\Gamma}_{z\bar{z}\bar{z}z}$ is the dominant component of the effective pairing interaction $\widetilde{\Gamma}(\bm{q})$. Its secondary peaks, indicated by the wave vector $\bm{Q}_{2}$ originating from FS nesting between the $\alpha$ and $\beta$ pockets [see Fig.~\ref{FS_band}(b)], also determine that the gap functions on the $\alpha$ and $\beta$ pockets have opposite signs. The $\alpha$ pocket is mainly associated with the bonding state of the $d_{x^{2}-y^{2}}$ orbital [Fig.~\ref{FS_band}(b)]. The pairing functions $\widetilde{\phi}_{\bar{z}\bar{z}}$ and $\widetilde{\phi}_{xx}$ follow this relation:
\begin{align}
\lambda \widetilde{\phi}_{\bar{z}\bar{z}}(\bm{k}) \sim& -\frac{T}{N}\sum_{\bm{q}} \widetilde{\Gamma}_{z\bar{z}\bar{z}z}(\bm{q})\widetilde{G}_{zx}(\bm{k}-\bm{q}) \widetilde{G}_{zx}(\bm{q}-\bm{k}) \nonumber \\
&\times \widetilde{\phi}_{xx}(\bm{k}-\bm{q}).
\end{align}
Thus, $\widetilde{\phi}_{\bar{z}\bar{z}}$ and $\widetilde{\phi}_{xx}$ must satisfy $\widetilde{\phi}_{\bar{z}\bar{z}}(\bm{k})\widetilde{\phi}_{xx}(\bm{k}-\bm{Q}_{2})<0$, which results in the bonding state of the $d_{x^{2}-y^{2}}$ orbital and the antibonding state of the $d_{z^{2}}$ orbital, linked by $\bm{Q}_{2}$ [Fig.~\ref{layer_s}(d)], having opposite signs [Figs.~\ref{bond_s}(a) and (d)]. Consequently, the gap functions on the $\alpha$ and $\beta$ pockets also display opposite signs.

We also find that in the BA representation, the spin fluctuations $\widetilde{\chi}^s_{z\bar{z}z\bar{z}}$ and $\widetilde{\chi}^s_{x\bar{x}x\bar{x}}$ dominate the effective pairing interaction $\widetilde{\Gamma}_{z\bar{z}\bar{z}z}$ among various collective fluctuations. According to Eqs.~(\ref{gamma-q-eqx}) and~(\ref{U-O-BA}), their contributions to $\widetilde{\Gamma}_{z\bar{z}\bar{z}z}$ can be expressed as:
\begin{align}
    \widetilde{\Gamma}_{z\bar{z}\bar{z}z} \sim U^2\widetilde{\chi}^s_{z\bar{z}z\bar{z}} + J^2\widetilde{\chi}^s_{x\bar{x}x\bar{x}}.
\end{align}
As illustrated in Figs.~\ref{susceptibility}(a) and (b) in Appendix~\ref{app}, $\widetilde{\chi}^s_{z\bar{z}z\bar{z}}$ and $\widetilde{\chi}^s_{x\bar{x}x\bar{x}}$ are responsible for the two peaks of $\widetilde{\Gamma}_{z\bar{z}\bar{z}z}$ at $\bm{Q}_{1}$ and $\bm{Q}_{2}$, respectively.

\subsection{The case for $V_z=0.5$ eV}

\begin{figure}
    \centering
    \includegraphics[width=\columnwidth]{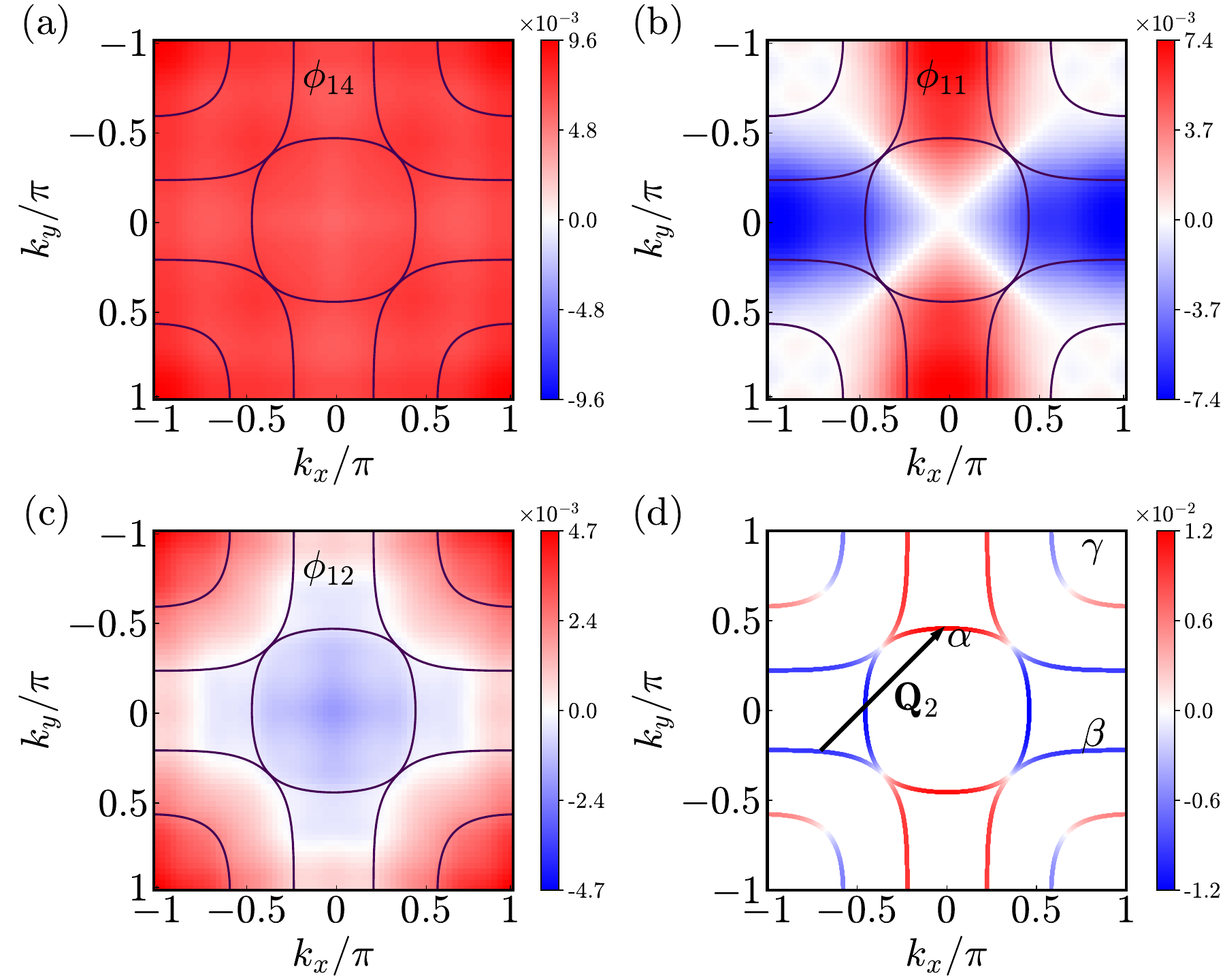}
    \caption{Pairing functions for $U=1.0$ eV and $V_z=0.5$ eV in the orbital-layer representation. (a) Interlayer pairing $\phi_{14}$ between the $d_{x^{2}-y^{2}}$ and $d_{z^{2}}$ orbitals. (b) Intralayer pairing $\phi_{11}$ for the $d_{x^2-y^2}$ orbital. (c) Intralayer pairing $\phi_{12}$ between the $d_{x^{2}-y^{2}}$ and $d_{z^{2}}$ orbitals. (d) Gap Function projected onto the Fermi surface, with black arrow $\boldsymbol{Q}_2$ indicating sign change.}
    \label{layer_d}
\end{figure}
Next, we explore how the interlayer interaction $V _{z}$ affects the superconducting pairing symmetry, with a particular focus on the results for $U=1.0$ eV and $V_z=0.5$ eV. Figures~\ref{layer_d}(a)-(c) illustrate the three main components of $\phi(\bm{k})$, which differ from the primary components observed when $V_z=0$ [see Figs.~\ref{layer_s}(a)-(c)]. The interlayer pairing $\phi_{14}$ between the $d_{x^2-y^2}$ and $d_{z^2}$ orbitals, along with the intralayer pairing $\phi_{11}$ of the $d_{x^2-y^2}$, become the primary pairing components. As shown in Fig.~\ref{layer_d}(d), the gap function projected onto the FS exhibits a $d_{x^2-y^2}$-wave symmetry,  contrasting with the $s_{\pm}$-wave symmetry observed when $V_{z}=0$. This pairing functions can be approximately expressed as:
\begin{align}
\phi_{ij}(k)\!=\!
\left\{
\begin{array}{ll}
    \!\Delta_{\alpha} & ij\!=\!14~\text{or}~ 23, \\
    \!\Delta_{\beta1}(\cos k_x\!-\!\cos k_y)\\
    \!+\Delta_{\beta2}(\cos2k_x\!-\!\cos2k_y) & ij\!=\!11~\text{or}~33, \\
    \!0 & \text{otherwise},
\end{array}
\right.
\label{eq:d-wave}
\end{align}
where $\Delta_{\alpha}:\Delta_{\beta1}:\Delta_{\beta2}=1:-0.48:-0.15$ and $\phi_{ij}(k)=\phi_{ji}(k)$. Although the symmetries of the pairing functions in this case are entirely different from those for $V_{z}=0$, they still satisfy constraint condition (\ref{eq:constraint}). In addition, this $d$-wave superconductivity is dominated by interlayer pairing between the $d_{x^2-y^2}$ and $d_{z^2}$ orbitals, which significantly differs from the $d$-wave superconductivity dominated by intralayer pairing within the $d_{x^2-y^2}$ orbital in cuprate superconductors. The interorbital and interlayer characteristics of this $d$-wave superconductivity also differ markedly from the $d$-wave superconductivity based on the $d_{x^2-y^2}$ single-orbital model for La$_3$Ni$_2$O$_7$~\cite{PhysRevB.110.024514,PhysRevLett.132.146002}, which exhibits a single-layer $d$-wave pairing.

\begin{figure}
    \centering
    \includegraphics[width=\columnwidth]{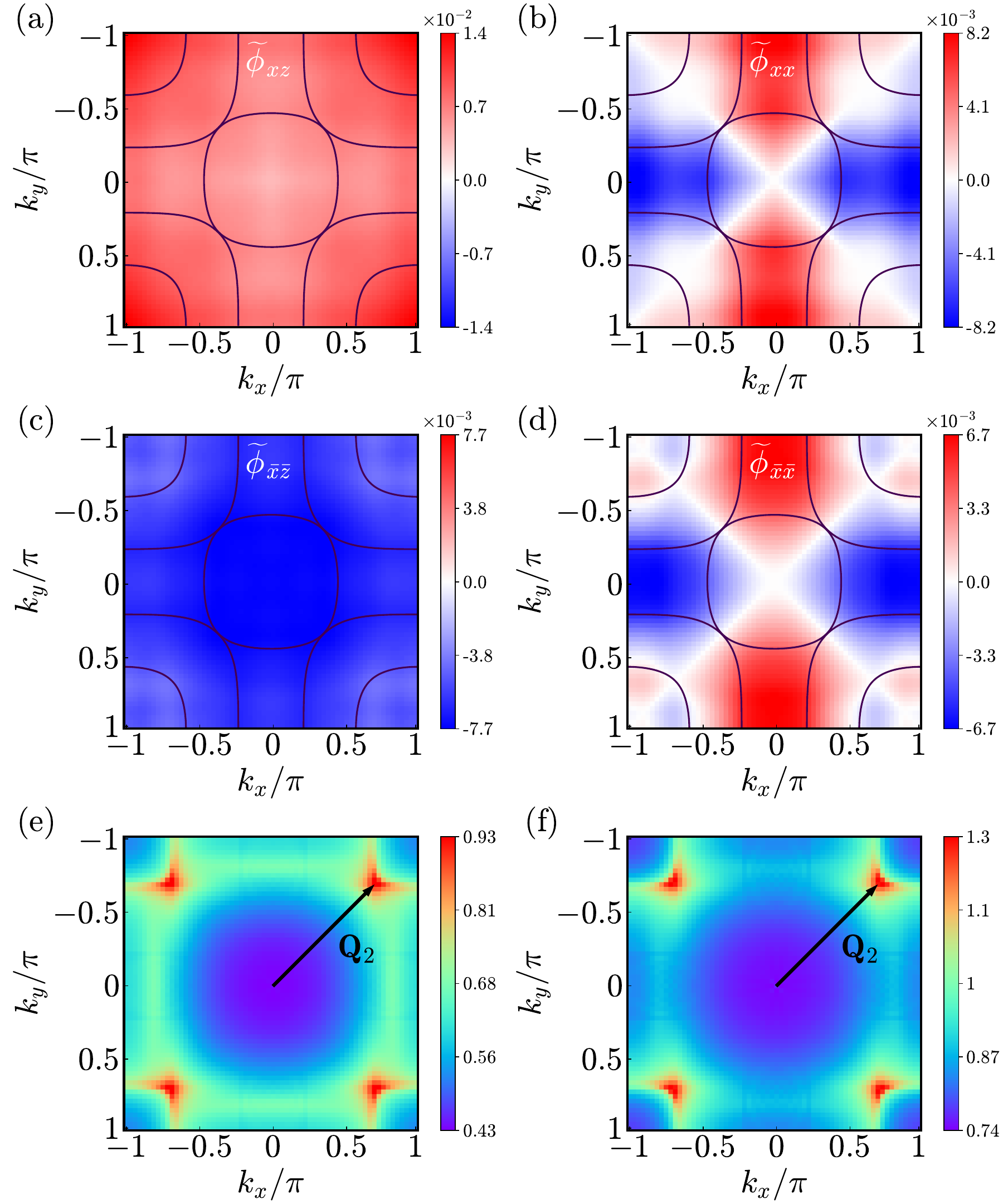}
    \caption{Pairing function and pairing interaction in the bonding-antibonding representation for $U=1.0$ eV and $V_z=0.5$ eV. (a)-(d) Pairing functions. (e) Pairing interaction $\widetilde{\Gamma}_{x\bar{x}\bar{z}z}(\bm{k})$. (f) Pairing interaction $\widetilde{\Gamma}_{x\bar{x}\bar{x}x}(\bm{k})$. The black arrows $\bm{Q}_2$ in (e) and (f) indicating peaks.}
    \label{bond_d}
\end{figure}
Using the same approach applied to the $s_{\pm}$-wave superconducting state at $V_{z}=0$,  we employ the BA representation to investigate the origin of $d_{x^2-y^2}$-wave state at  $V_z=0.5$ eV. In Figs.~\ref{bond_d}(a)-(d), we display the main components of the pairing function $\widetilde{\phi}(\bm{k})$. We find that, for both the bonding and antibonding states, the dominant pairing is interorbital, which contrasts sharply with the $V_z=0$ situation, where the intraorbital pairings are predominant [Figs.~\ref{bond_s}(a)-(d)]. Notably, the intraorbital pairing of the $d_{z^{2}}$ orbital is no longer favorable due to the interlayer repulsive interaction $V_{z}$. Furthermore, we find that the pairing functions $\widetilde{\phi}_{xz}$ and $\widetilde{\phi}_{\bar{x}\bar{z}}$ for bonding and antibonding states, respectively, have opposite signs. This can be explained by the following relation:
\begin{align}
\lambda \widetilde{\phi}_{\bar{x}\bar{z}}(\bm{k}) \sim& -\frac{T}{N}\sum_{\bm{q}} \widetilde{\Gamma}_{x\bar{x}\bar{z}z}(\bm{q})\widetilde{G}_{zx}(\bm{k}-\bm{q})
\widetilde{G}_{xz}(\bm{q}-\bm{k}) \nonumber \\
&\times \widetilde{\phi}_{xz}(\bm{k}-\bm{q}).
\end{align}
As illustrated in the Fig.~\ref{bond_d}(e), the interorbital pairing interaction $\widetilde{\Gamma}_{x\bar{x}\bar{z}z}$ peaks at the wave vector $\bm{Q}_2$. This results in $\widetilde{\phi}_{xz}$ and $\widetilde{\phi}_{\bar{x}\bar{z}}$  satisfying $\widetilde{\phi}_{xz}(\bm{k})\widetilde{\phi}_{\bar{x}\bar{z}}(\bm{k}-\bm{Q}_2)<0$, leading to the opposite signs of $\widetilde{\phi}_{xz}$ and $\widetilde{\phi}_{\bar{x}\bar{z}}$. Similarly, the intraorbital pairing functions $\widetilde{\phi}_{xx}$ and $\widetilde{\phi}_{\bar{x}\bar{x}}$ satisfy the following relation:
\begin{align}
    \lambda \widetilde{\phi}_{\bar{x}\bar{x}}(\bm{k}) \sim &-\frac{T}{N}\sum_{\bm{q}} \widetilde{\Gamma}_{x\bar{x}\bar{x}x}(\bm{q})\widetilde{G}_{xx}(\bm{k}-\bm{q})\widetilde{G}_{xx}(\bm{q}-\bm{k}) \nonumber \\
    &\times \widetilde{\phi}_{xx}(\bm{k}-\bm{q}).
\end{align}
As depicted in the Fig.~\ref{bond_d}(f), $ \widetilde{\Gamma}_{x\bar{x}\bar{x}x}$ has a similar structure to $\widetilde{\Gamma}_{x\bar{x}\bar{z}z}$, resulting in opposite signs for the bonding-state pairing $\widetilde{\phi}_{xx}$  and the antibonding-state pairing $\widetilde{\phi}_{\bar{x}\bar{x}}$, linked by $\bm{Q}_2$ [Figs.~\ref{bond_d}(b) and (d)]. These sign changes between the bonding-state and antibonding-state pairings collectively result in the gap function exhibiting $d$-wave symmetry, as shown in Fig.~\ref{layer_d}(d).

Similar to the case when $V=0$, for $V_z = 0.5$ eV, the effective pairing interactions are also primarily driven by spin fluctuations. The two key effective pairing interactions $\widetilde{\Gamma}_{x\bar{x}\bar{z}z}$ and $\widetilde{\Gamma}_{x\bar{x}\bar{x}x}$ discussed above are primarily governed by the spin fluctuations  $\widetilde{\chi}^s_{\bar{x}x\bar{z}z}$ and $\widetilde{\chi}^s_{x\bar{x}x\bar{x}}$ [see Figs.~\ref{susceptibility}(c) and (d) in Appendix~\ref{app}], respectively. According to Eqs.~(\ref{gamma-q-eqx}) and~(\ref{U-O-BA}), the relationships between them are:
\begin{align}
    \widetilde{\Gamma}_{x\bar{x}\bar{z}z}\sim U(U+V_z)\widetilde{\chi}^s_{\bar{x}x\bar{z}z}
\end{align}
and
\begin{align}
    \widetilde{\Gamma}_{x\bar{x}\bar{x}x}\sim U^2\widetilde{\chi}^s_{x\bar{x}x\bar{x}}.
\end{align}

\subsection{Evolution of pairing symmetry with interactions}

\begin{figure}
    \centering
    \includegraphics[width=\columnwidth]{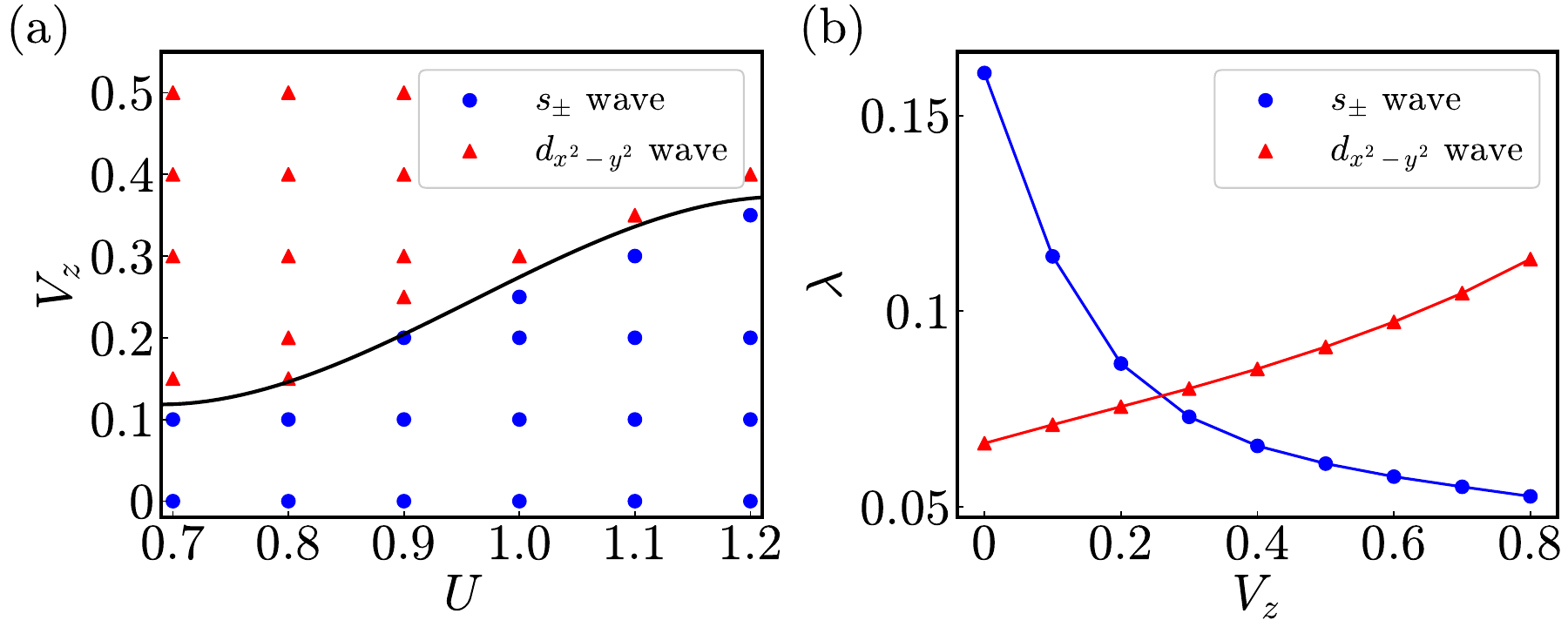}
    \caption{(a) $U$-$V_{z}$ phase diagram of pairing symmetry. The solid line represents the critical value for the superconducting transition. (b) Dependence of $\lambda$ on $V_z$ for $s_{\pm}$-wave and $d_{x^2-y^2}$-wave pairings at $U = 1.0$ eV.}
    \label{phase}
\end{figure}
We then examine how the pairing symmetry evolves with variations in the on-site Hubbard interaction $U$ and the interlayer Coulomb interaction $V_{z}$. The phase diagram for pairing symmetry in the $(U,V_{z})$ parameter space, obtained using the FLEX method, is shown in Fig.~\ref{phase}(a). For small values of $V_{z}$, the pairing symmetry is $s_{\pm}$-wave. However, as $V_{z}$  increases, the  $d_{x^{2}-y^{2}}$-wave superconducting state becomes more favorable. The critical value of $V_{z}$ for the transition from the $s_{\pm}$-wave state to the $d_{x^{2}-y^{2}}$-wave state increases with $U$.

\begin{figure}
    \centering
    \includegraphics[width=\columnwidth]{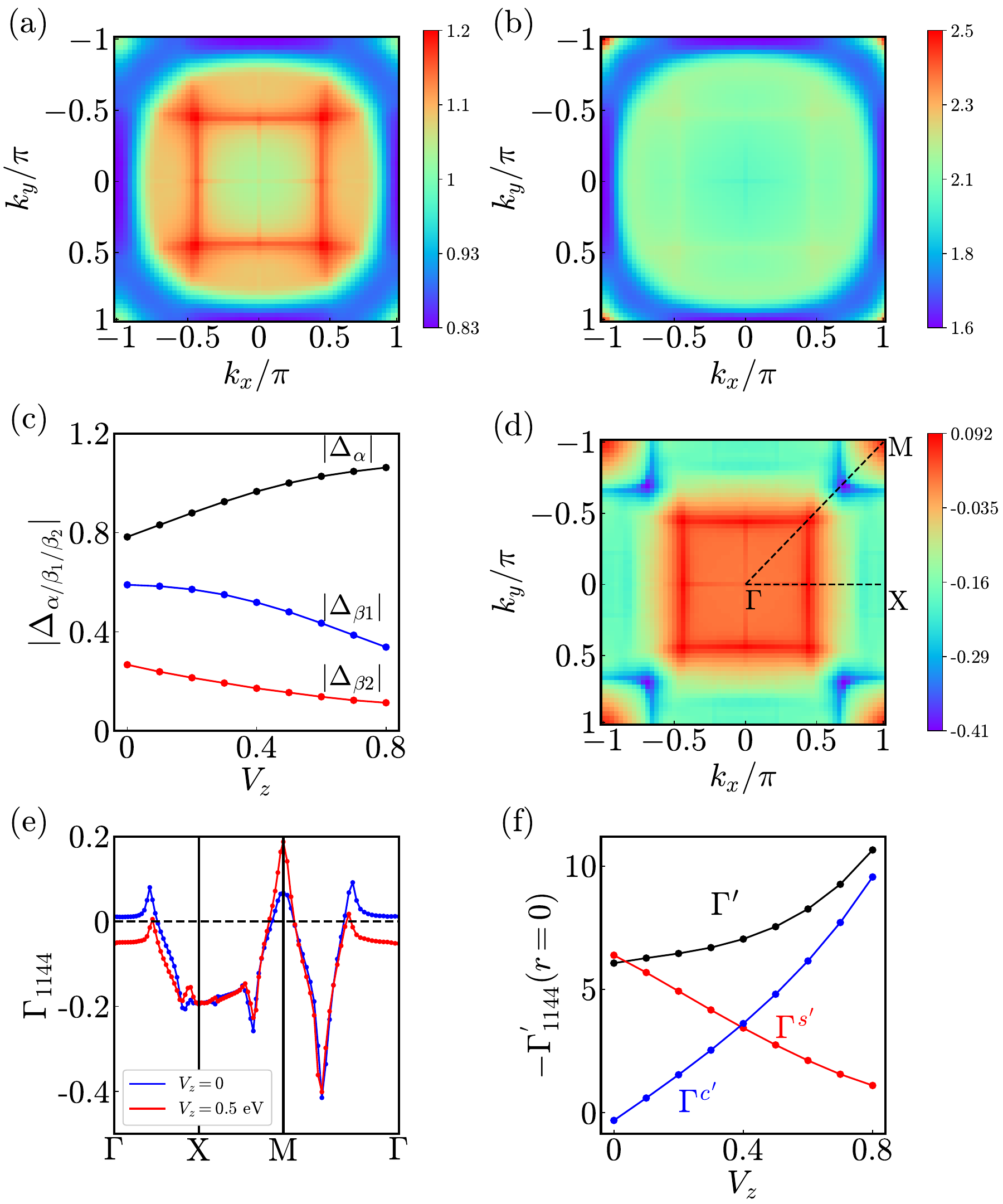}
    \caption{ (a) Pairing interaction $\widetilde{\Gamma}_{zzzz}$ for $V_z=0$. (b) $\widetilde{\Gamma}_{zzzz}$ for $V_z=0.5$ eV; (c) Evolution of the pairing amplitudes in Eq.~(\ref{eq:d-wave}) as a function of $V_{z}$. (d) Pairing interaction $\Gamma_{1144}$ for $V_z=0$. (e) Intensities of the pairing interaction $\Gamma_{1144}$ along high-symmetry lines shown in (d) for both $V_z=0$ and $V_z=0.5$ eV. (f) Evolution of the pairing potential $\Gamma^{\prime}_{1144}(\boldsymbol{r}=0)$ and the contributions of the spin and charge fluctuations with varying $V_z$.}
    \label{sus_change}
\end{figure}

As illustrated in Fig.~\ref{phase}(b), the eigenvalue $\lambda$ in Eq.~(\ref{gap-eq}) for the $s_{\pm}$-wave state decreases with increasing $V_z$, whereas the eigenvalue for the $d_{x^2-y^2}$-wave state increases. With the rise of $V_z$, the component $\widetilde{\Gamma}_{zzzz}$ of pairing interaction $\widetilde{\Gamma}(\bm{q})$ changes significantly more than the other components. As depicted in Figs.~\ref{sus_change}(a) and (b), the introduction of $V_z$ leads to a substantial increase in $\widetilde{\Gamma}_{zzzz}$, with its peak shifting to $\bm{Q}_{3}=(\pi, \pi)$. Considering the following relation:
\begin{align}
\lambda \widetilde{\phi}_{zz}(\bm{k}) \sim &- \frac{T}{N}\sum_{\bm{q}} \widetilde{\Gamma}_{zzzz}(\bm{q})\widetilde{G}_{zz}(\bm{k}-\bm{q})\widetilde{G}_{zz}(\bm{q}-\bm{k}) \nonumber \\
&\times \widetilde{\phi}_{zz}(\bm{k}-\bm{q}),
\end{align}
it is clear that $\widetilde{\Gamma}_{zzzz}$ tends to favor $\widetilde{\phi}_{zz}(\bm{k})\widetilde{\phi}_{zz}(\bm{k}-\bm{Q}_3)<0$, requiring $\widetilde{\phi}_{zz}$ to have a $d$-wave form. In contrast, the constrain condition (\ref{eq:constraint}) restricts $\widetilde{\phi}_{zz}$ to an $s$-wave form for the $s_{\pm}$-wave state. Consequently, the increase of $V_{z}$ suppresses the $s_{\pm}$-wave state. This aligns with the observation in real space that the interlayer interaction $V_{z}$ can suppress the primary $d_{z^2}$-orbital interlayer pairing in Eq.~(\ref{gap-spm-x1}).

To understand why an increase in $V_z$ benefits the $d_{x^2-y^2}$-wave state, we analyze how the three pairing amplitudes in Eq.~(\ref{eq:d-wave}) evolve with $V_z$. As shown in Fig.~\ref{sus_change}(c), only the amplitude $\Delta_{\alpha}$ increases, while the others decrease. This suggests that the pairing components $\phi_{14}$  and $\phi_{41}$ become increasingly dominant as $V_z$ rises. Thus, to understand how $V_z$ enhances the $d_{x^2-y^2}$-wave state, we can examine its effect on $\phi_{14}$ and $\phi_{41}$ using the following relation in the orbital-layer representation:
\begin{align}
    \lambda \phi_{14}(\boldsymbol{k}) \sim& -\frac{T}{N}\sum_{\boldsymbol{q}}\Gamma_{1144}(\boldsymbol{q})G_{14}(\boldsymbol{k}-\boldsymbol{q})
        G_{41}(\boldsymbol{q}-\boldsymbol{k}) \nonumber \\
        &\times \phi_{41}(\boldsymbol{k}-\boldsymbol{q}).
        \label{v1144}
\end{align}
The interorbital pairing interaction $\Gamma_{1144}$ between the $d_{x^2-y^2}$ and $d_{z^2}$ orbitals on different layers for $V_z=0$ is depicted in Fig.~\ref{sus_change}(d). A key feature of $\Gamma_{1144}$ is its predominance of negative values. This generally requires $\phi_{14}$ to maintain its sign across the entire BZ, characteristic of the $d_{x^2-y^2}$-wave state [see Fig.~\ref{layer_d}(a)]. As illustrated in Fig.~\ref{sus_change}(e), with increasing $V_z$, the negative region of $\Gamma_{1144}$, which favors $d_{x^2-y^2}$-wave state, also expands. Thus, an increase in $V_z$ promotes the emergence of $d_{x^2-y^2}$-wave superconductivity. To clarify this further, we perform a Fourier transform on $\Gamma_{1144}(\bm{q})$ to obtain the effective pairing potential $\Gamma^{\prime}_{1144}(\bm{r})$ in real space. We find that $\Gamma^{\prime}_{1144}(\bm{r}=0)$ acts as an attractive interaction for two electrons on the same site, consistent with the $s$-wave form of $\phi_{14}$. In Fig.~\ref{sus_change}(f), we demonstrate the evolution of $\Gamma^{\prime}_{1144}(\bm{r}=0)$ along with the contributions from spin and charge fluctuations ($\Gamma^{\prime s}(\bm{r}=0)$ and $\Gamma^{\prime c}(\bm{r}=0)$) as a function of $V_z$. The functions $\Gamma^{\prime s}(\bm{r})$ and $\Gamma^{\prime c}(\bm{r})$ are obtained via Fourier transformations of $\Gamma^{s}_{1144}(\bm{q})$ and $\Gamma^{c}_{1144}(\bm{q})$ as described in Eq.~(\ref{gamma-q-eqx}). It is observed that the charge-fluctuation contribution $\Gamma^{\prime c}(\bm{r}=0)$ becomes dominant as $V_z$ increases. Specially, according to Eq.~(\ref{gamma-q-eqx}), the effects of $V_z$ on $\Gamma^{c}_{1144}$ can be simplified as:
\begin{align}
\Gamma^c_{1144}\sim -U^c_{1122}\chi^c_{2222}U^c_{2244} = -2V_z(2U'-J)\chi^c_{2222}.
\end{align}
We find that the charge fluctuation of $d_{z^2}$ orbital ($\chi^c_{2222}$) plays a crucial role in affecting the evolution of $\Gamma^{c}_{1144}$ with $V_z$ and consequently plays a crucial role in the realization of $d_{x^2-y^2}$-wave superconductivity.

\section{\label{summary}Summary}

Motivated by the strong interlayer coupling in the bilayer nickelate La$_{3}$Ni$_{2}$O$_{7}$, we  investigated the impact of interlayer Coulomb interactions on superconducting pairing symmetry. Using the fluctuation-exchange approximation on a bilayer two-orbital model with $d_{x^2-y^2}$ and $d_{z^2}$ orbitals, we found that the interlayer Coulomb interaction $V_{z}$ between the $d_{z^2}$ orbitals can induce a change in the superconducting gap from $s_{\pm}$-wave symmetry to $d_{x^2-y^2}$-wave symmetry. In the $s_{\pm}$-wave superconducting state with small  $V_{z}$, the dominant pairing is intraorbital within the $d_{z^2}$ orbital, exhibiting similar amplitudes for both intralayer and interlayer components. Conversely, in the $d_{x^2-y^2}$-wave superconducting state with large $V_{z}$, the leading pairing involves interlayer interactions between the $d_{x^2-y^2}$ and $d_{z^2}$ orbitals. In the $s_{\pm}$-wave superconducting state, intraorbital pairing components exhibit $s$-wave symmetry, while interorbital components exhibit $d$-wave symmetry. In contrast, in the $d_{x^2-y^2}$-wave superconducting state, intraorbital components show $d$-wave symmetry, and interorbital components show $s$-wave symmetry.  Additionally, the mirror symmetry of the bilayer structure allows us to describe pairing symmetries using a bonding-antibonding basis. In this context, the $s_{\pm}$-wave superconductivity primarily involves intraorbital pairing within the antibonding $d_{z^2}$ orbital, whereas the $d_{x^2-y^2}$-wave superconductivity  is mainly characterized by interorbital pairing between the bonding $d_{x^2-y^2}$ and $d_{z^2}$ orbitals.

\begin{acknowledgements}
This work was supported by National Key Projects for Research and Development of China (No. 2024YFA1408104 and No. 2021YFA1400400) and the National Natural Science Foundation of China (No. 12074175, No. 92165205, No. 12374137, and No. 12434005).
\end{acknowledgements}

\appendix

\section{\label{app}Spin susceptibilities}

\begin{figure}
    \centering
    \includegraphics[width=\columnwidth]{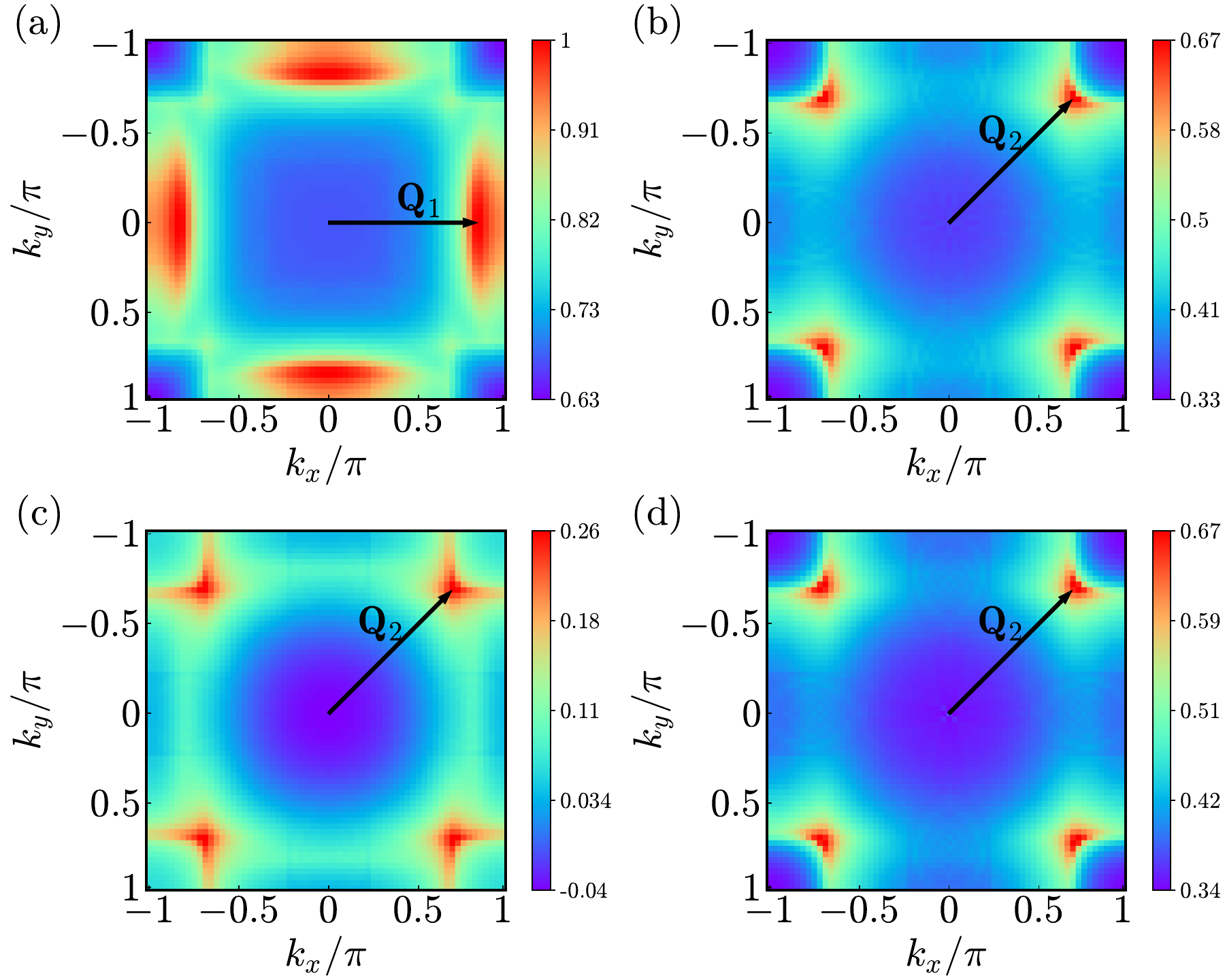}
    \caption{Susceptibilities in the BA representation. (a) $\widetilde{\chi}^s_{z\bar{z}z\bar{z}}$ for $V_z=0$. (b) $\widetilde{\chi}^s_{x\bar{x}x\bar{x}}$ for $V_z=0$. (c) $\widetilde{\chi}^s_{\bar{x}x\bar{z}z}$ for $V_z=0.5$ eV (d) $\widetilde{\chi}^s_{x\bar{x}x\bar{x}}$ for $V_z=0.5$ eV.}
    \label{susceptibility}
\end{figure}
In the FLEX approximation, the effective pairing interactions stem from various collective fluctuations. In the model we explore in this paper, spin fluctuations play a dominant role. We present the spin susceptibilities $\widetilde{\chi}^s_{z\bar{z}z\bar{z}}$ and $\widetilde{\chi}^s_{x\bar{x}x\bar{x}}$ in Figs.~\ref{susceptibility}(a) and (b), which primarily contribute to the effective interaction $\widetilde{\Gamma}_{z\bar{z}\bar{z}z}$ at $V_z=0$. Similarly, the spin susceptibilities $\widetilde{\chi}^s_{\bar{x}x\bar{z}z}$ and $\widetilde{\chi}^s_{x\bar{x}x\bar{x}}$ shown in Figs.~\ref{susceptibility}(c) and (d) are key contributors to the effective interactions $\widetilde{\Gamma}_{x\bar{x}\bar{z}z}$ and $\widetilde{\Gamma}_{x\bar{x}\bar{x}x}$ at $V_z=0.5$ eV. Here, $\widetilde{\chi}^s$ represents the spin susceptibility in the BA representation, which can be derived from $\chi^{s}$ in the orbital-layer representation using the transformation matrix $U$ in Eq.~(\ref{U-O-BA}). The peaks in these susceptibilities, indicated by $\bm{Q}_{1}$ and $\bm{Q}_{2}$, originate from the nesting properties between the bonding and antibonding pockets of FS, as shown in Fig.~\ref{FS_band}(b).

\bibliography{ref}

\end{document}